\documentclass[aps,pre,twocolumn,showpacs,amsmath]{revtex4}

\usepackage{bm}
\usepackage{graphicx}
\usepackage{hyperref}
\begin{document}
\title{Partly melted DNA conformations obtained with a probability peak finding method}
\author{Eivind T\o stesen}
\email[Correspondence: ]{eivindto@radium.uio.no}
\homepage[Web server: ]{http://stitchprofiles.uio.no/}
\affiliation{Department of Tumor Biology, The Norwegian Radium Hospital,
N-0310 Oslo, Norway}
\date{\today}

\begin{abstract}
Peaks in the probabilities of loops or bubbles, helical segments,
and unzipping ends in melting DNA are found in this article using
a peak finding method that maps the hierarchical structure of certain energy
landscapes. The peaks indicate the alternative conformations
that coexist in equilibrium and the range of their fluctuations. This yields
a representation
of the conformational ensemble at a given temperature, which is illustrated
in a single diagram called a stitch profile.
This article describes the methodology and discusses
stitch profiles vs.\ the ordinary
probability profiles using the phage lambda genome as an example.
\end{abstract}

\maketitle

\section{\label{intro}Introduction}
DNA melts by a stochastic formation and growth of loops\footnote{Loops
are here synonymous to \emph{bubbles} and should be distinguished from
looping in which the helix as a whole bends back on itself.}
and tails (i.e.\ unzipping ends). Loop formation is also induced in
the dense cellular environment
and is at the heart of DNA biology \cite{sumner}.
For the last three decades,
numerically calculated properties of the DNA melting process have been
represented
by plotting curves of three types: probability profiles, melting curves,
and
Tm profiles. Probability profiles \cite{poland,Yerafizz} are plots of the
basepairing probability
$p_{\text{bp}}(i)$ or the ``upside-down'' $1-p_{\text{bp}}(i)$
vs.\  sequence position $i$.
Melting curves \cite{BD98,meltsim} are plots of the helicity $\Theta$ or
its derivative vs.\
temperature $T$.
Tm profiles are plots of the basepair melting temperature
$T_{\text{m}}(i)$ vs.\  sequence position $i$. Apparently, there has been
less interest in
calculating other properties, one reason being perhaps that
the ordinary plots outline
the main features on the experimental side, such
as the melting curves from UV spectroscopy and
differential scanning calorimetry. However, there
are other interesting properties within reach of
calculations. For example, what are the sizes and
locations of loops and how do they
fluctuate? How do distant loops correlate? What
are the alternative conformations of
a region that coexist when it melts? What events are
predominant and what others are rare? In addition
to these questions being important
\emph{per se}, advances in single molecule techniques
\cite{altan,danilo,zeng}
provide new types of measurement of the micromechanical, dynamical,
and structural
properties, and motivate predictions beyond the ordinary curves
\cite{hanke,hwa,peyrardreview}.

\begin{figure}
 \includegraphics[height=6.3cm]{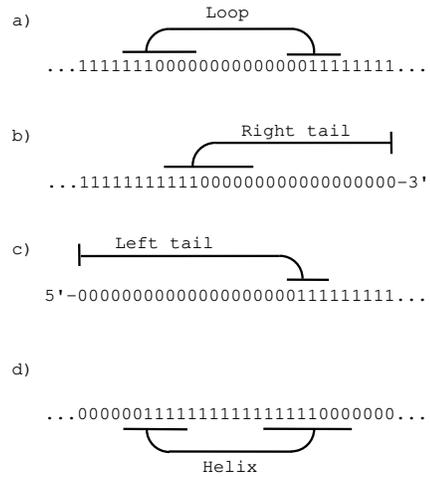}%
 \caption{
 A stitch profile is composed of four types of stitches: (a) loops,
 (b) right tails,
 (c) left tails, and (d) helices. Loop and tail stitches are drawn
 on the upper side and they span regions of opened base-pairs (0's).
 Helix stitches are drawn on the lower side and they span regions
 of closed base-pairs (1's). The horizontal bars indicate fluctuational
 ranges of
 the 0-1 boundaries.
 \label{stitchesfig}
}
 \end{figure}

In this article, we turn our attention to stitch profiles.
A \emph{stitch profile} is a
diagrammatic representation of the alternative DNA conformations
that coexist at a given temperature \cite{bip2003}.
Figure \ref{stitchesfig}
shows the four types of graphical elements called \emph{stitches}
that go into a stitch profile. A stitch represents either a loop, a
right tail,
a left tail or a helical region, as shown, and indicates its boundary
positions and
the ranges of fluctuation of these positions. In analogy with sewing,
where a thread
coming up and down through the fabric forms a row of stitches, any
conformation of
DNA is an alternating row of blocks of open or closed basepairs. A
stitch profile
indicates alternative conformations as alternative
threads (i.e. paths) through the diagram.
The aim of this work is to
develop a method for constructing stitch profiles and to discuss
stitch profiles vs. probability profiles.

In the Poland-Scheraga model \cite{PS}, a state of the chain molecule
is specified by $N$ binary variables, $x_{1},\ldots,x_{N}$, where the
$j$-th variable
$x_{j}$ ($= 0 \text{ or } 1$)
indicates if the $j$-th base in the sequence is basepaired or not with
the complementary
strand. While the classical three types of curves are based on
calculating the basepairing probabilities related to the state $x_{j}$
of each basepair,
a stitch profile, in contrast, is based on probabilities of blocks of
basepairs
being in states corresponding to loops, helical segments or tails.
A stitch profile made ``by hand'' was introduced in Ref.~\onlinecite{bip2003}
(where we referred to it as a loop map) in order to suggest an application
of
such block probabilities. The article described a
DNA melting algorithm with two important
features: a speedup based on multiplication of
symmetrical leftside and rightside
partition functions; and the statistical weight
of a basepair depending rigorously on
both of its neighbors. These features allow
the block probabilities to be easily
calculated as follows.

A loop is a consecutive series of 0's (melted basepairs)
bounded by 1's at positions
$x$ and $y$, where $1\leq x<y-1<N$. The probability
of a loop is calculated by decomposing the
chain in three segments,
\begin{equation}
p_{\text{loop}}(x,y)=Z_{\text{X10}}(x)\Omega(y-x)Z_{\text{01X}}(y)/\beta Z.
\label{ploop}
\end{equation}
$Z_{\text{X10}}(x)$ is a partition function characterizing
the segment $[1,x+1]$, $Z_{\text{01X}}(y)$ is a partition function
characterizing the segment $[y-1,N]$, $Z$ is the total
partition function of the whole chain,
$\Omega(y-x)$ is the
loop entropy factor (a function of loop size), and $\beta$
is related to the equilibrium constant
of complete dissociation of the two strands. A tail is a
block of 0's that extends to the end of the chain. The probability
of a right tail from position $N$ (the right chain end) to a bounding 1
at position $x$ is
\begin{equation}
p_{\text{right}}(x)=Z_{\text{X10}}(x)/\beta Z.
\label{pright}
\end{equation}
The probability of a left tail from position 1 (the left chain end) to a
bounding 1 at position $y$
is
\begin{equation}
p_{\text{left}}(y)=Z_{\text{01X}}(y)/\beta Z.
\label{pleft}
\end{equation}
A helical region is a block of 1's bounded by 0's or
by the chain ends. If $x$ is the position of
the first 1 in the block and $y$ is the position
of the last, where $1\leq x\leq y\leq N$,
then the probability can be written
\begin{equation}
p_{\text{helix}}(x,y)=Z_{\text{X01}}(x)\Xi(x,y)Z_{\text{10X}}(y)/\beta Z,
\label{phelix}
\end{equation}
where $Z_{\text{X01}}(x)$ is a partition function characterizing the
segment $[1,x]$ and $Z_{\text{10X}}(y)$ is a
partition function characterizing
the segment $[y,N]$. The \emph{stacking chain}
function $\Xi(x,y)$ is the statistical
weight of the block of 1's, given as
\begin{equation}
\Xi(x,y)=\left\{
                     \begin{array}{l}
                     s_{1}(x)s_{1}(y)\prod_{j=x+1}^{y}s_{11}(j)
		                    \text{ for } x<y \\
                     s_{010}(x)
		                    \text{ for } x=y,
		     \end{array}
		\right.
\label{xi}
\end{equation}
where $s_{11}(x)$ is the statistical weight of
nearest neighbor basepairs (a pair of 1's),
$s_{1}(x)$ is the statistical weight of a
helix-ending basepair (0 on one side, 1 on the other),
and $s_{010}(x)$ is the statistical weight
of an isolated basepair (0's on both sides).
Eqs.~(\ref{ploop})--(\ref{phelix}) correspond
to Eqs.~(12)--(15) in Ref.~\onlinecite{bip2003},
respectively.

The block probabilities $p_{\text{loop}}$,
$p_{\text{right}}$, $p_{\text{left}}$,
and $p_{\text{helix}}$ depend on precisely
located boundaries $x$ and/or $y$. But
thermal motion causes the boundaries to
fluctuate. These fluctuations are represented
by fluctuation bars in a stitch profile.
They are not merely attributes like
"error bars", but rather an essential ingredient.
Each stitch represents not a single
conformation of a region, but a grouping of
conformations that are supposed to be related via
fluctuations. In a plot of any of the block
probabilities as a function of
$x$ and/or $y$, such a grouping will appear
as a broad peak, and the fluctuation bar(s) indicate
the extent of the peak. A stitch profile is
simply a representation of the peaks in the
four block probability functions, and the
problem of constructing a stitch profile is
basically a peak finding problem.

The peak finding problem is important in
data and signal analysis in diverse
areas of science, for example, in various
types of spectroscopy and image analysis.
Peak finding has also been used in
statistical mechanics to define macroscopic
states of RNA secondary structure: native,
intermediate, molten, and denatured states
\cite{TCD2001}.
The following issues apply to our case here:
A peak's size is given by its
volume rather than its height. If a peak is
broad enough, it may have a higher volume than
another peak, even if it has a lower height.
Probability peak heights can be very low,
so we can not use a height cutoff for
detecting peaks. There is no erroneous noise in our
calculated probabilities, so smoothing
should not be used. Actually, the shapes of peaks are
quite irregular with ``peaks within peaks'',
so the problem is \emph{hierarchical} peak finding
(analogous to hierarchical clustering vs. clustering).
There may be limited space in a stitch
profile, so it must be chosen which peaks to represent.

The right and left tail stitches are found by
1D peak finding in $p_{\text{right}}$
and $p_{\text{left}}$, respectively,
and the loop and helix stitches are found by
2D peak finding in $p_{\text{loop}}$
and $p_{\text{helix}}$, respectively. The
challenge is not so much finding a peak as
deciding its extent. In 1D we use an interval---the
fluctuation bar---to delimit a peak. In 2D
we use a \emph{frame}, that is, the cartesian
product of
the two fluctuation bars on the $x$-axis and
the $y$-axis, to represent a peak by ``framing'' it.
Let the \emph{peak volume} $p_{\text{v}}$ be
defined by the probability summed over the interval in 1D or
the frame in 2D.

This article describes a probability peak finding method in 1D
based on a detailed mapping of the hierarchical structure. The 2D case is
solved by combining the 1D results for $x$ and $y$. For the
extent of a peak, the main idea is to find where
the probability has dropped to a certain fraction relative to the
peak maximum value. This fraction
is controlled by a parameter to the algorithm
and it determines the widths of the fluctuation bars.
These widths, in turn, determine the peak volumes that can
be used for choosing if stitches are included or not in the stitch profile.
\section{\label{methods}The Method}
Minus the logarithm of a probability is an energylike quantity.
Using this, we transform probability peak finding into
finding the wells or lakes in a (pseudo)energy landscape.
A peak with a certain ratio between the probability at the maximum and
the probabilities at the edges corresponds to a lake with a certain depth in
an energy landscape. The analogy to mountain landscapes,
lakes, ponds, etc., is standard in statistical mechanics \cite{stein}.
We use it here to redefine the stitch profile problem of
peak finding to be a lake finding problem in
four energy landscapes---two 1D landscapes:
\begin{eqnarray}
E_{1}(x) &=& -\log_{10}p_{\text{right}}(x),
\\
E_{2}(y) &=& -\log_{10}p_{\text{left}}(y),
\end{eqnarray}
and two 2D landscapes:
\begin{eqnarray}
E_{3}(x,y) &=& -\log_{10}p_{\text{loop}}(x,y),
\\
E_{4}(x,y) &=& -\log_{10}p_{\text{helix}}(x,y).
\end{eqnarray}
\subsection{Peak finding method in 1D}
\begin{figure*}
 \includegraphics[height=9.0cm]{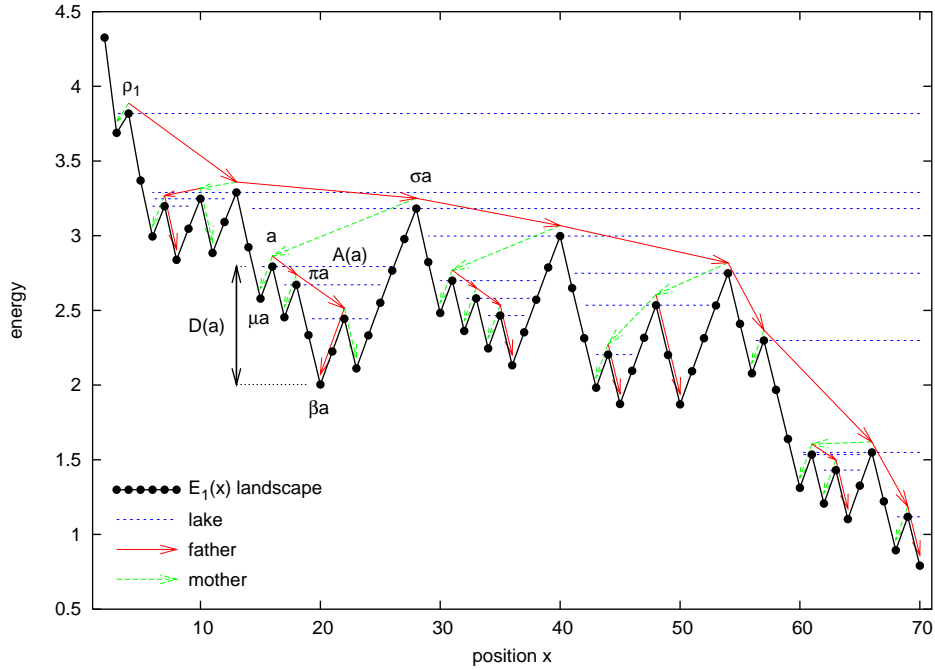}%
 \caption{
(Color online) $E_{1}(x)$ is plotted for a 70 bp sequence
to illustrate the pedigree ordering of lakes in an energy
landscape. Lakes corresponding to each local
maximum are shown as horizontal dashed
(blue) lines.
Arrows indicate the binary tree starting from the root $\rho_{1}=4$.
Paternal lines are connected series of fathers shown as solid (red) arrows.
The node $a=16$, for example, has the
lake $L(a)=[15,26]$, bottom $\beta a=20$, depth
$D(a)\simeq 0.8$, father $\pi a=18$,
mother $\mu a=15$, and successor $\sigma a=28$, as shown.
 \label{landfig}
}
 \end{figure*}

The 1D method is described here using $E_{1}(x)$ as an example,
but $E_{2}(y)$ is treated the same way.
Of all the possible lakes
that can be created by filling water into the
various wells, we restrict ourselves to considering only
a finite set of representative lakes.
Let $\Psi_{1}$ be the set of sequence positions $a$
at which $E_{1}$ has an extremum.
Minima and maxima in $\Psi_{1}$ are alternating along the $x$-axis.
To each element  $a\in \Psi_{1}$ we
associate a \emph{lake} $L(a)$ in the landscape, see Fig.~\ref{landfig}.
The altitude of the surface (water level) is $E_{1}(a)$
and the lake surface spans the interval
$L(a)=[L_{\text{L}}(a),L_{\text{R}}(a)]$ given by
\begin{subequations}
\label{lake}
\begin{eqnarray}
& a\in L(a) \label{lakea},
\\
& \forall x \in L(a):E_{1}(x)\leq E_{1}(a) \label{lakeb},
\\
& E_{1}(L_{\text{L}}(a)-1)>E_{1}(a) \text{ or }
 L_{\text{L}}(a)=1 \label{lakec},
\\
& E_{1}(L_{\text{R}}(a)+1)>E_{1}(a) \text{ or }
 L_{\text{R}}(a)=N \label{laked}.
\end{eqnarray}
\end{subequations}
When $a$ is a minimum, in most cases $L(a)=\{a\}$.
When $a$ is a maximum, the lake $L(a)$ is nearly split in
two adjacent lakes,
divided at position $a$ where the local depth becomes zero.
However the corresponding probability peak is not split in
two by a zero probability, so we consider $L(a)$ as one lake.
The \emph{bottom} $\beta a$ of a lake
$L(a)$ is defined as the position with lowest energy
in the lake,
\begin{equation}
\beta a=\arg \min_{x \in L(a)} E_{1}(x)
\end{equation}
The \emph{depth} $D(a)$ of a lake $L(a)$ is defined
as the energy difference between the surface and the
bottom: $D(a)= E_{1}(a)- E_{1}(\beta a)$. Some lakes are
contained inside deeper lakes:  $L(a)\subset L(b)$. This
partial ordering of lakes in $\Psi_{1}$ defines a
hierarchical structure \cite{hoffmann}.
Assuming that both the leftmost and the rightmost
(on the $x$-axis) elements in $\Psi_{1}$
are minima (terminal
maxima could just be excluded), the elements
in $\Psi_{1}$ can be considered as the nodes of a binary tree.
The root $\rho_{1}$ of the tree is the global maximum
and its lake $L(\rho_{1})$ spans the entire sequence (or almost).
\subsubsection{Pedigree ordering}
Imagine a walk along the branches of the binary tree.
In order to orient itself, the walker needs
``roadsigns'' at each node $a$ that point the directions
to the root $\rho_{1}$ and the bottom $\beta a$.
This imposes a structure similar to a pedigree
(i.e.\ tree of ancestors) as illustrated in
Fig.~\ref{landfig}. Each node $a\neq \rho_{1}$ is connected
upwards in the direction of the root to a unique node
$\sigma a$ called the \emph{successor} of $a$. This means
$L(a)\subset L(\sigma a)$.
Each node $a$ corresponding to a maximum is also connected
downwards to two parent nodes: a \emph{father} node $\pi a$ in the
direction of the bottom; and a \emph{mother} node $\mu a$
in the other direction.
This means that the father is the parent with the lowest
bottom, $E_{1}(\beta \pi a)<E_{1}(\beta \mu a)$, and that
$\beta \pi a=\beta a$.
But it does \emph{not} imply that $D(\pi a)>D(\mu a)$.
In contrast to an offspring tree, parents are located
in the direction away from the root, rather than the reverse.
Define the set of successors of a node $a$:
\begin{equation}
\Sigma (a)=\{a,\sigma a,\sigma ^{2}a,\sigma ^{3}a,\ldots,\rho _{1}\}.
\end{equation}
The set $\Sigma (a)$ traces a path from $a$ up to the root.
Define the set of ancestors of a node $a$:
\begin{equation}
\Delta (a)=\{b\in \Psi_{1}|a\in \Sigma (b)\}.
\end{equation}
The set $\Delta (a)$ is the subtree that has $a$ as its root or top node.
Each node $a\in \Psi_{1}$ belongs to a unique \emph{paternal line},
\begin{equation}
\Pi (a)=\{\varphi a,\ldots,a,\pi a,\ldots,\beta a\},
\end{equation}
which is a maximal set of nodes that are related through a
series of fathers. The series ends at their common bottom node $\beta a$
and, oppositely,  it begins at a node,
called the \emph{full} node $\varphi a$,
which is not itself a father.
Each node corresponding to a minimum is
the bottom of its paternal line. And
each node which is either a mother or the root $\rho_{1}$
is the full node of its paternal line. For a node $a$ that
is both a minimum and a mother, $\Pi (a)=\{a\}$.
The term ``full''
stems from filling water into a bottom: The
paternal line indicates successively deeper lakes
until the full lake is reached. The successor
of the full lake belongs to another paternal line
and corresponds to another bottom being
filled.
\subsubsection{The \textsc{maxdeep} algorithm}
\begin{figure}
 \includegraphics[height=6.3cm]{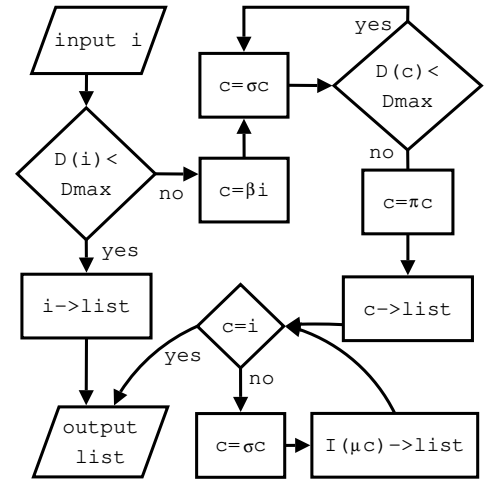}%
 \caption{
 The \textsc{maxdeep} algorithm that
 finds stitches with depths below $D_{\text{max}}$.
 For a given input node $i$, it returns a list of those nodes in
 $\Delta (i)$ that fulfill the criterion in Eq.~(\ref{crit}).
 $c$ is the tree climber, $\rightarrow$
 means push (i.e.\ ``put on a list''),
 and $I(\mu c)$ is the output of a recursive call
 to the algorithm itself with $\mu c$
 as the input node. The diagram can be
 read in conjunction with Fig.~\ref{landfig}.
 \label{Dmaxfig}
}
 \end{figure}

The lake finding task at hand is to search through
the set of lakes corresponding to the set $\Psi_{1}$,
and select the lakes to be represented in a stitch profile.
A possible solution is the \textsc{maxdeep} algorithm.
With a given parameter $D_{\text{max}}$,
the algorithm finds all nodes $a\in \Psi_{1}$ where
\begin{subequations}
\label{crit}
\begin{eqnarray}
D(a) &<& D_{\text{max}}
\\
D(\sigma a) &\geq & D_{\text{max}} \text{ or } a=\rho _{1}.
\end{eqnarray}
\end{subequations}
In words, this means that they should be as deep as
possible without exceeding the maximum depth.
In many cases, it can be expected that the depth
increases in just a small step from a node to its
successor, but not for full nodes that often have
successors that are much deeper than themselves. Therefore,
some of the selected lakes will have depths close
to $D_{\text{max}}$, while others (the full ones) can
be more shallow. The \textsc{maxdeep} algorithm
is shown in Fig.~\ref{Dmaxfig}.
It is not necessary to evaluate all $a\in \Psi_{1}$.
Instead, the \textsc{maxdeep} algorithm involves a
``tree climber'' $c$ that basically climbs up the
paternal line $\Pi (\rho _{1})$ of the input node
$i=\rho _{1}$, starting from the bottom $\beta \rho _{1}$,
until it exceeds $D_{\text{max}}$, and then takes
one step down again. At that point, $c$ fulfills
the criterion in Eq.~(\ref{crit}), while
all other nodes in $\Sigma (c)$ and $\Delta (c)$ do not.
Subsequently, the algorithm calls itself
recursively with mothers of $\Sigma (c)$
as input nodes, in order
to explore other paternal lines.

\begin{figure}
 \includegraphics[width=8.6cm]{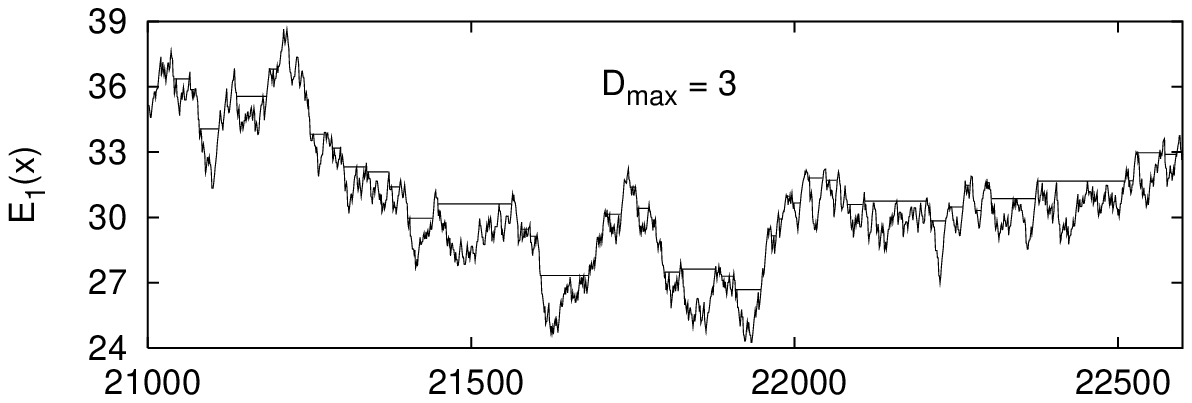}%

 \includegraphics[width=8.6cm]{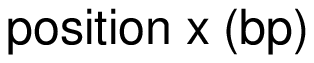}%

 \includegraphics[width=8.6cm]{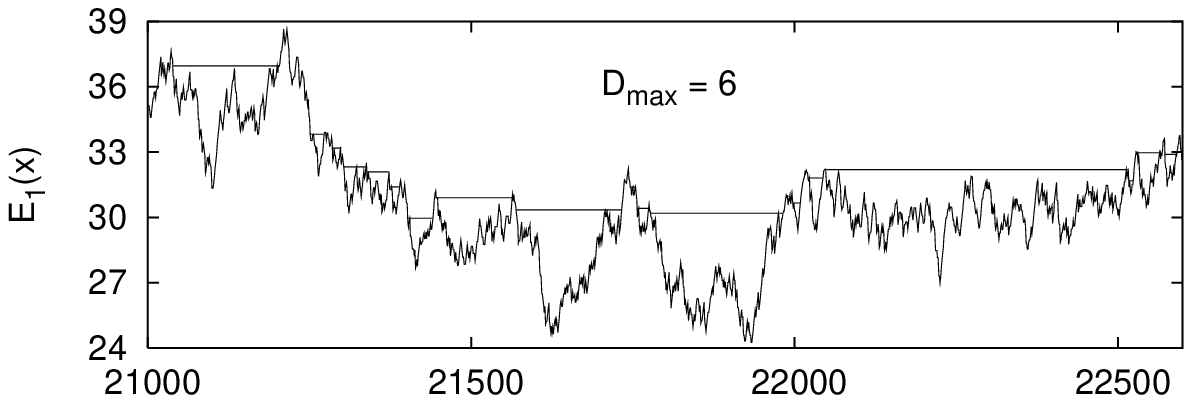}%

 \includegraphics[width=8.6cm]{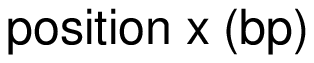}%
 \caption{
Lakes in 1D found by the \textsc{maxdeep} algorithm.
The two plots show the same region of an energy landscape. The first plot
shows the lakes found by the algorithm using $D_{\text{max}}=3$ and
the second plot shows the lakes with $D_{\text{max}}=6$.
 \label{lakesfig}
}
 \end{figure}

The output of the \textsc{maxdeep} algorithm
is illustrated in Fig.~\ref{lakesfig}.
Lakes have depths between zero and $D_{\text{max}}$ and they cover
a large fraction of the landscape.
The figure shows that the effect of increasing $D_{\text{max}}$
is that some lakes become wider and deeper,
some lakes merge, and some lakes
(corresponding to full nodes) remain unchanged.
\subsubsection{The largest peaks}
Some of the lakes found by the \textsc{maxdeep} algorithm
are probably not very significant, having low
depths and low peak volumes at the same time.
A second selection process is
required before we finally get the stitches for the profile.
Since the \textsc{maxdeep} algorithm does not
consider the peak volumes $p_{\text{v}}$,
these can be considered in the second selection.
If we want to select the largest peaks, for example, this can be achieved
by having a \emph{cutoff} as a second parameter  $p_{\text{c}}$
and select only nodes $a$ with peak volume
$p_{\text{v}}(a)\geq p_{\text{c}}$.
\subsection{Peak finding method in 2D}
While a 1D lake is completely described as
an interval $[L_{\text{L}}(a),L_{\text{R}}(a)]$,
lakes in the 2D landscapes $E_{3}$ and $E_{4}$ have complicated
contours and perhaps islands in the interior \cite{hoffmann}.
Fortunately, we only need
to know a 2D lake's extents on the $x$-axis and the $y$-axis,
in order to enclose it in a frame, which is
needed in a stitch profile (cf. section \ref{intro}).
\subsubsection{The helix landscape}
Consider the lake finding problem in the helix landscape $E_{4}$.
Although isolated basepairs are allowed
[$x=y$ in Eq.~(\ref{phelix})] we will ignore
those instances for convenience, and require
$1\leq x< y\leq N$ in the landscape. It follows from
Eqs.~(\ref{phelix}) and (\ref{xi}) that
\begin{equation}
p_{\text{helix}}(1,y)p_{\text{helix}}(x,N)=
p_{\text{helix}}(x,y)p_{\text{helix}}(1,N).
\end{equation}
Defining two 1D landscapes,
\begin{eqnarray}
E_{5}(x) &=& -\log_{10}p_{\text{helix}}(x,N),
\\
E_{6}(y) &=& -\log_{10}p_{\text{helix}}(1,y),
\end{eqnarray}
we can write $E_{4}(x,y)=E_{5}(x)+E_{6}(y)+\text{const}$.
This decoupling of $x$ and $y$ allows us to
analyze the lakes in $E_{4}$, based on 1D
analysis of $E_{5}$ and $E_{6}$ and
their binary trees $\Psi_{5}$ and $\Psi_{6}$.

Let $a\in \Psi_{5}$ and $b\in \Psi_{6}$ and consider
the frame $L(a)\times L(b)$. Does such a frame
enclose a 2D lake in $E_{4}$? At least, the frame should
cover a region of $E_{4}$.
The frame is said to be \emph{above the
diagonal} if $L_{\text{R}}(a)<L_{\text{L}}(b)$,
such that all $(x,y)\in L(a)\times L(b)$
are points in the helix landscape. Because of the
decoupling, the minimum in a frame is
\begin{equation}
\arg \min_{(x,y)\in L(a)\times L(b)} E_{4}(x,y)=(\beta a,\beta b).
\end{equation}
Consider the energy barriers seen from
$(\beta a,\beta b)$:
\begin{eqnarray}
\Delta E_{4}(x,y) &=& E_{4}(x,y)-E_{4}(\beta a,\beta b)
\nonumber \\
 &=& E_{5}(x)-E_{5}(\beta a)+E_{6}(y)-E_{6}(\beta b)
\nonumber \\
 &=& \Delta E_{5}(x)+\Delta E_{6}(y).
\end{eqnarray}
Using Eq.~(\ref{lakeb}), we find that
$\Delta E_{4}(x,\beta b)\leq D(a)$ for all $x\in L(a)$, and
using Eqs.~(\ref{lakec}) and (\ref{laked}), we find that
$\Delta E_{4}(x,y)> D(a)$ just outside the west and east side
of the frame.
This means that $L(a)$ is the extent on the $x$-axis
of a 2D lake with depth $D(a)$ and bottom $(\beta a,\beta b)$.
And vice versa, $L(b)$ is the extent on the $y$-axis
of a 2D lake with depth $D(b)$ and bottom $(\beta a,\beta b)$.
If $D(a)\neq D(b)$, then a 2D lake with  bottom $(\beta a,\beta b)$
must have depth $D\leq\min \{D(a),D(b)\}$
to be confined by the frame, and it will not extend to all four
frame sides. A lake with depth
$D>\min \{D(a),D(b)\}$ is not confined and may not even have
its bottom inside the frame.
If $D(a)=D(b)$ then the frame  $L(a)\times L(b)$
is exactly the extent in both dimensions
of a 2D lake with that depth.

Unfortunately, we can not expect to
find $a$'s and $b$'s with equal depths.
In order to best approximate 2D lakes, we want instead
$D(a)$ and $D(b)$ to be as close as possible, that is,
none of $\pi a$, $\sigma a$, $\pi b$ or $\sigma b$ should have
depths in between $D(a)$ and $D(b)$. This can be formulated as
two conditions: we say that $(a,b)$ is ``$\sigma$-above'' if
\begin{subequations}
\label{1}
\begin{eqnarray}
D(\sigma b)>D(a) \text{ or } b=\rho _{6},
\\
D(\sigma a)>D(b) \text{ or } a=\rho _{5},
\end{eqnarray}
\end{subequations}
and we say that $(a,b)$ is ``$\pi$-below'' if
\begin{subequations}
\label{2}
\begin{eqnarray}
D(\pi b)<D(a) \text{ or } b=\beta b,
\\
D(\pi a)<D(b) \text{ or } a=\beta a.
\end{eqnarray}
\end{subequations}
Frames that are $\sigma$-above, $\pi$-below, and above the diagonal
are good representations of lakes with depth $\min \{D(a),D(b)\}$.
Let us examine the three conditions one by one.
First, define the set $\Gamma_{4}$ of frames that are $\sigma$-above,
\begin{equation}
\Gamma_{4}=\{(a,b)\in \Psi_{5}\times \Psi_{6}|
(a,b)\text{ is }\sigma \text{-above}\}.
\end{equation}
Eq.~(\ref{1}) shows that $(\rho _{5},\rho _{6})\in \Gamma_{4}$.
And $(a,b)\in \Gamma_{4}$ if $a$ and $b$ both correspond to minima,
because Eq.~(\ref{1}) is true with
$D(a)=0$ and $D(b)=0$.
This means $(\beta a,\beta b)\in \Gamma_{4}$ for any
$a\in \Psi_{5}$ and
$b\in \Psi_{6}$.
Just as 1D lakes are hierarchically ordered, so are frames. A frame is
contained in another frame, $L(a)\times L(b)\subset L(c)\times L(d)$,
if $a\in \Delta (c)$ and $b\in \Delta (d)$. It turns out that
the elements of $\Gamma_{4}$ are the nodes of a binary tree with
$(\rho _{5},\rho _{6})$ as its root.
We call $\Gamma_{4}$ the \emph{frame tree},
and it is a kind of \emph{product tree}
between the trees $\Psi_{5}$ and $\Psi_{6}$.
The successor of a node $(a,b)\neq (\rho _{5},\rho _{6})$ is
\begin{equation}
\sigma(a,b)=\left\{
 \begin{array}{l}
  (\sigma a,b) \text{ if } D(\sigma b)>D(\sigma a) \text{ or } b=\rho_{6}\\
  (a,\sigma b) \text{ if } D(\sigma a)>D(\sigma b) \text{ or } a=\rho_{5}
 \end{array}
\right.
\label{sigma}
\end{equation}
Each node $(a,b)\in \Gamma_{4}$, where $a$ and $b$ are not both minima, has
two parent nodes.
We define the bottom of a node as $\beta (a,b)=(\beta a,\beta b)$,
which enables us to distinguish the two parent nodes as a father,
\begin{equation}
\pi(a,b)=\left\{
                     \begin{array}{l}
                     (\pi a,b) \text{ if } D(a)>D(b)\\
                     (a,\pi b) \text{ if } D(a)<D(b),
 		     \end{array}
		\right.
\label{pi}
\end{equation}
with $\beta \pi (a,b)=\beta (a,b)$, and a mother
\begin{equation}
\mu(a,b)=\left\{
                     \begin{array}{l}
                     (\mu a,b) \text{ if } D(a)>D(b)\\
                     (a,\mu b) \text{ if } D(a)<D(b),
 		     \end{array}
		\right.
\label{mu}
\end{equation}
with $\beta \mu (a,b)\neq \beta (a,b)$.
This gives the frame tree a pedigree ordering with paternal lines, etc.,
just as in 1D.

Some frames in $\Gamma_{4}$ are not above the diagonal,
such as the root frame $(\rho _{5},\rho _{6})$ that spans almost the
entire sequence in both dimensions.
Next, define the set of frames that are $\sigma$-above \emph{and}
above the diagonal:
\begin{equation}
\Psi_{4}=\{(a,b)\in \Gamma_{4}|L_{\text{R}}(a)<L_{\text{L}}(b)\}.
\end{equation}
$\Psi_{4}$ is organized as a number of
disjoint binary trees, $\Psi_{4}=\bigcup_{j}\Delta (a_{j},b_{j})$,
each one being a subtree of $\Gamma_{4}$. The top node
$(a_{j},b_{j})$ of the $j$-th subtree is above the diagonal, while
its successor $\sigma (a_{j},b_{j})$ crosses the diagonal.
And lastly, define the set of frames that are
$\sigma$-above, $\pi$-below, and above the diagonal:
\begin{equation}
\Upsilon_{4}=\{(a,b)\in \Psi_{4}|(a,b)\text{ is }\pi \text{-below}\}.
\end{equation}
This set also consists of disjoint trees, but they are not binary,
nodes can have more than two parents.

If we do not require $\pi$-below and consider the larger set
$\Psi_{4}$, then such frames are still good representations of lakes.
A computational advantage of this is that each subtree
$\Delta (a_{j},b_{j})$ in $\Psi_{4}$ can be searched
with the \textsc{maxdeep} algorithm, by
using its top node $(a_{j},b_{j})$ as the input $i$.
The $E_{4}$ lake finding problem is solved by finding
frames in $\Psi_{4}$ using the \textsc{maxdeep} algorithm in this manner,
followed by a second selection based on the cutoff $p_{\text{c}}$,
which yields the helix stitches for a stitch profile.
For the \textsc{maxdeep} algorithm we must define the depth of a frame.
A possible depth definition is
$D(a,b)=\max \{D(a),D(b)\}$. Using this,
the \textsc{maxdeep} algorithm finds
frames $(a,b)$ with $D(a)<D_{\text{max}}$ and  $D(b)<D_{\text{max}}$.
\subsubsection{The loop landscape}
Consider the lake finding problem in the loop landscape $E_{3}$.
It follows from Eqs.~(\ref{ploop})--(\ref{pleft}) that for
$1\leq x<y-1<N$,
\begin{equation}
E_{3}(x,y)=E_{1}(x)+E_{2}(y)-\log_{10}\Omega(y-x)+\text{const}.
\end{equation}
$x$ and $y$ do not decouple in $E_{3}$.
But $\log_{10}\Omega(y-x)$ varies slowly enough to be
considered constant as an approximation. With this assumption,
it turns out that an analysis parallel to the one for the helix
landscape gives reasonable stitch profiles. There are only minor differences
between the loop and helix cases.
For example, a loop frame is said to be \emph{above the
diagonal} if $L_{\text{R}}(a)+1<L_{\text{L}}(b)$. Here is a brief outline:
We define the frame tree
\begin{equation}
\Gamma_{3}=\{(a,b)\in \Psi_{1}\times \Psi_{2}|
(a,b)\text{ is }\sigma \text{-above}\}.
\end{equation}
The root of $\Gamma_{3}$ is $(\rho _{1},\rho _{2})$.
By replacing $\rho _{5}$ with $\rho _{1}$ and  $\rho _{6}$ with $\rho _{2}$
in Eqs.~(\ref{1}) and (\ref{sigma})--(\ref{mu}) we define $\sigma$-above,
successors, fathers, and mothers in $\Gamma_{3}$.
The set of $\sigma$-above frames that are also above the diagonal is
\begin{equation}
\Psi_{3}=\{(a,b)\in \Gamma_{3}|L_{\text{R}}(a)+1<L_{\text{L}}(b)\}.
\end{equation}
And we search $\Psi_{3}$ for frames using the
top nodes of the subtrees and the \textsc{maxdeep} algorithm.
Subsequently, the loop stitches are found by a
selection based on the cutoff $p_{\text{c}}$.
\subsection{Stitch profile data}
\begin{figure}
 \includegraphics[width=8.6cm]{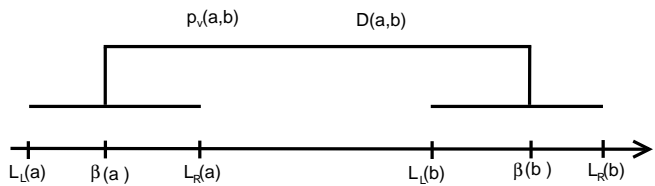}%
 \caption{
Eight quantities characterize a loop stitch.
 \label{row}
}
\end{figure}

A mere calculation of the block probabilities
[cf. Eqs.~(\ref{ploop})--(\ref{phelix})] gives
$O(N^{2})$ small numbers which are information
in a fragmented form.
In principle, these block probabilities
can be obtained as functions of $x$ and/or $y$
using the Poland-Scheraga model \cite{PS},
the Peyrard-Bishop model \cite{DPB93}, or something else.
The peak finding method is applied for putting this information
in the more useful form of a stitch profile,
which is data of size $O(N)$ only.
A stitch profile is a set of stitches of the four types
in Fig.~\ref{stitchesfig}. As we have seen, each fluctuation
bar corresponds to a lake in a 1D landscape. Figure \ref{row}
shows how the two fluctuation bars of a loop
stitch $(a,b)$ span the lake intervals $L(a)$ and $L(b)$, and
how the diagram also indicates the position of the lake bottom
$(\beta a,\beta b)$, where the probability peak has its
maximum. Two additional quantities are associated with a
stitch: the depth $D(a,b)$ and the peak volume $p_{\text{v}}(a,b)$.
These quantities can also be illustrated, for example, by labeling
the stitch. Thus, eight quantities are associated with
each loop or helix stitch, but only five quantities for
left or right tail stitches that only have one fluctuation bar.
\section{Discussion}
This section discusses different aspects of stitch profiles,
what information they represent, and the choice of
parameters and algorithm.
The 48 kbp phage lambda genome (GenBank accession number NC\_001416)
is used as a test sequence, to illustrate stitch profiles
and probability profiles rather than the melting behavior
vs.\ biology of lambda \cite{wada}.
All stitch profiles were calculated for the whole 48 kbp sequence,
but the interesting features are viewed in windows of length 1 kbp--20 kbp.
The stitch profiles in full length
are better viewed on a computer than in print
\footnote{See EPAPS Document No. ? for
an online view of the stitch profiles of lambda DNA at different temperatures}.
Partition functions were calculated
in the Poland-Scheraga model using the algorithm in Ref.~\onlinecite{bip2003}
with $\beta =1$. The parameter set of Blake \& Delcourt \cite{BD98}
at $[\text{Na}^{+}]=0.075\text{ M}$ was applied,
with the loop entropy factor $\Omega(y-x)=\sigma [2(y-x)+1]^{-\alpha }$
reparametrized by Blossey \& Carlon \cite{BC03}
using $\alpha =2.15$ and $\sigma =1.26\times 10^{4}$.
\subsection{Alternative conformations}
\begin{figure}
 \includegraphics[width=8.6cm]{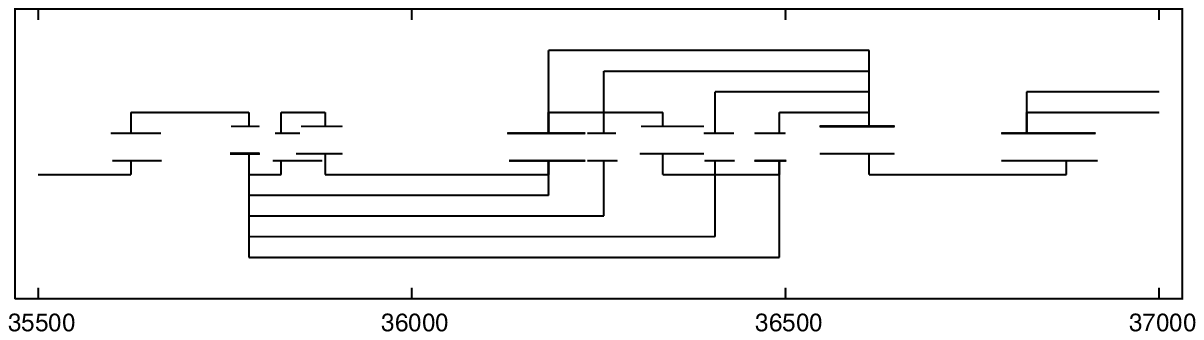}%

 \includegraphics[width=8.6cm]{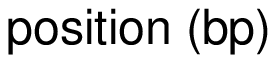}%

 \includegraphics[width=8.0cm]{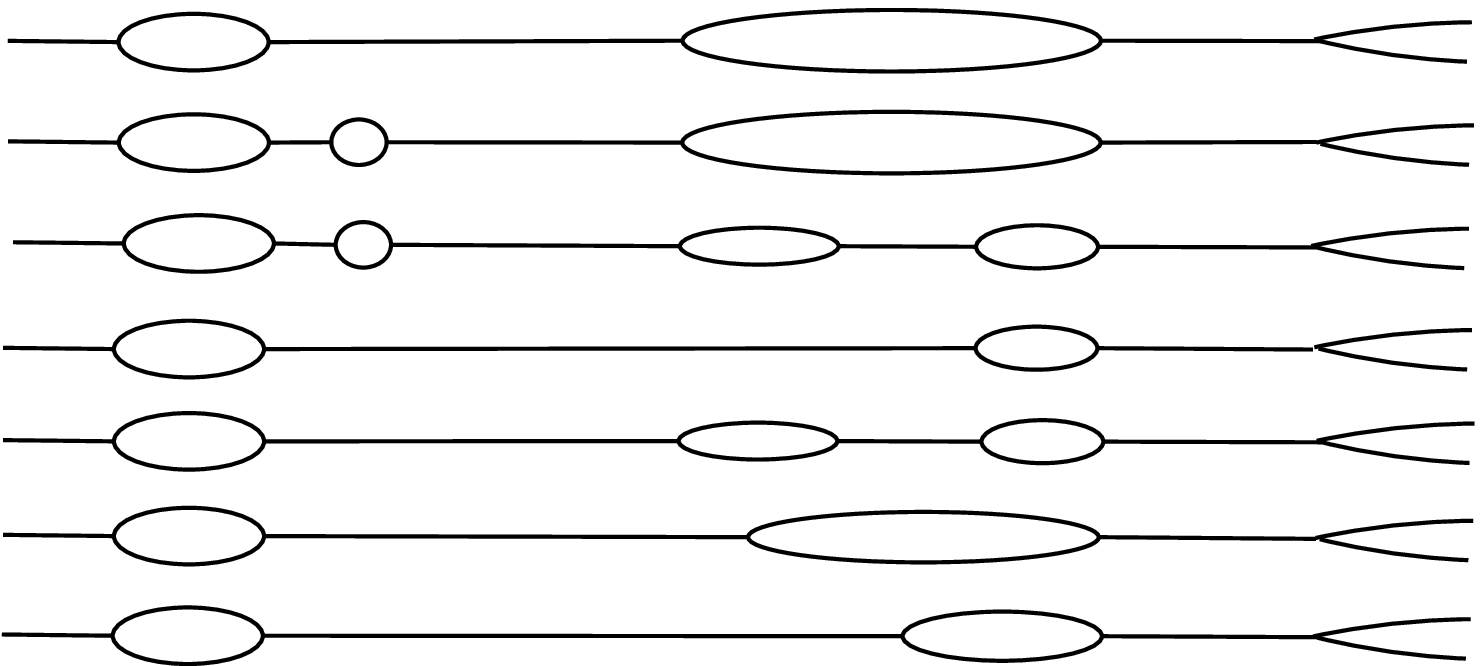}%
 \caption{
 A stitch profile (in the box) represents a number of alternative
 conformations. In this example, there are seven possible
 conformations listed schematically below the box.
 The parameters are $D_{\text{max}}=3$, $p_{\text{c}}=0.02$
 and  $T=81.9^{\circ }\text{C}$ and the sequence window is 35.5 kbp--37 kbp.
 \label{altconf}
 }
\end{figure}

As previously stated, a stitch profile shows a
number of alternative conformations
that coexist in equilibrium at a given temperature.
It does so in two ways:
(1) each stitch represents a fluctuational
variation of its boundaries,
and (2) alternative rows of stitches represent
alternative series of loops and helices along the chain.
Figure \ref{altconf}
shows a short sequence window of a stitch profile,
in which there are nine loop stitches and nine helix stitches.
Note how the fluctuation bars and the lake bottoms
of loops often coincide with those of helices, although
they were calculated independently of each other.
This is expected because a helix must begin where a
loop or tail ends, of course, so coinciding fluctuation bars
reflect the same boundary fluctuation.
A \emph{row of stitches} is a series of stitches connected
in a chain by coinciding boundaries,
and in the diagram, it forms a continuous path or thread,
which alternates between the upper and lower side.
There are often several stitches to choose among at a given boundary,
resulting in a combinatorial number of alternative rows of stitches.
Each row of stitches corresponds to
a specific conformation (apart from the fluctuational
variation) of a region that is much longer than the regions specified
by the individual stitches.
The stitch profile in Fig.~\ref{altconf} is aligned with a
schematic list of the seven alternative conformations corresponding
to the possible rows of stitches. These alternatives do not represent
the only possible conformations in that window---strictly speaking, any
conformation has a non-zero probability---but they represent the
most stable conformations in terms of probability peak volume and depth.

Note that Fig.~\ref{altconf} also illustrates
that stitches are sorted and stacked
vertically in the diagram according to their lengths $\beta b-\beta a$.
This is for aesthetic reasons only. There is no quantity associated with
the vertical axis. Fluctuation bars are also placed on different levels to
avoid overlap.
\subsection{Correlations and cooperativity}
\begin{figure*}
 \includegraphics[width=16cm]{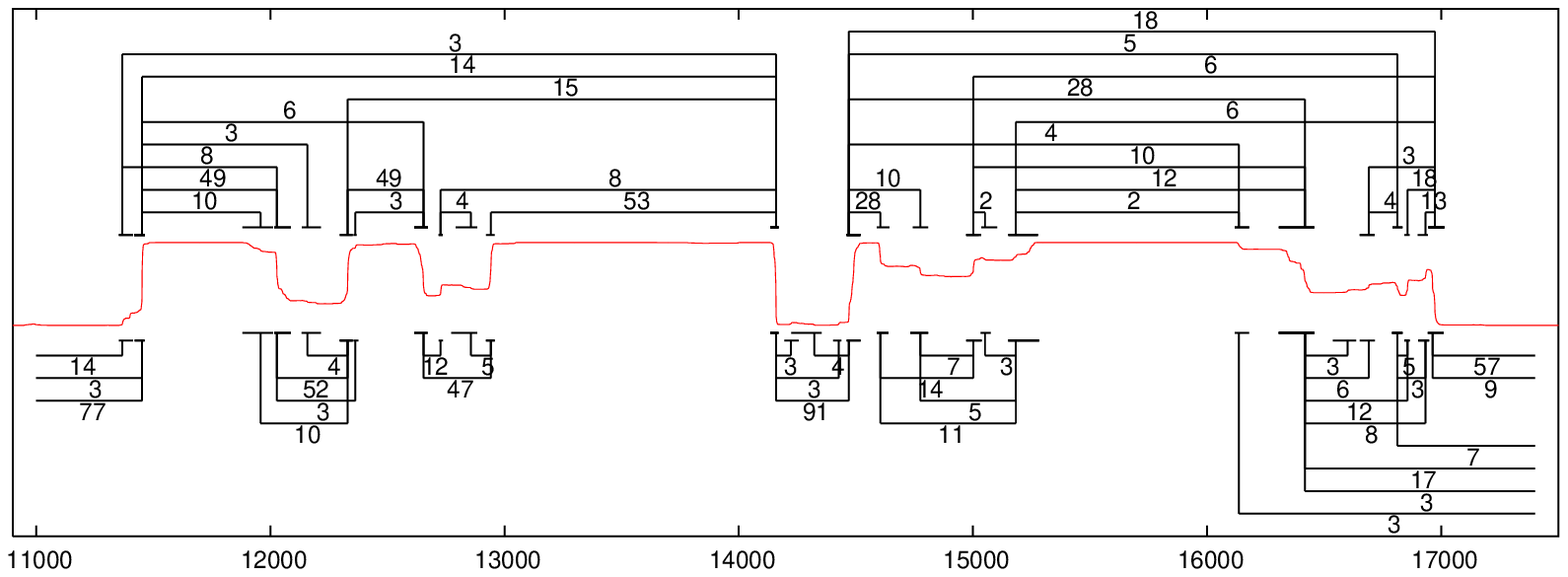}%

 \includegraphics[width=16cm]{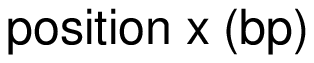}%
 \caption{
 (Color online) Comparison of a stitch profile and a probability profile,
 both calculated at $T\simeq 90^{\circ }\text{C}$
 where the helicity is $\Theta =0.1$.
 The curve in the middle (in red)
 is the probability profile $1-p_{\text{bp}}(x)$
 and it varies between 0 and 1 (vertical axis not shown).
 Each stitch is labeled with its peak volume $p_{\text{v}}$ in percent.
 The parameters are $D_{\text{max}}=3$ and $p_{\text{c}}=0.02$
 and the sequence window is 11 kbp--17.4 kbp.
 \label{prob}
 }
\end{figure*}

DNA cooperativity \cite{PS} is the presence of certain
long-range \emph{correlations}, that should be distinguished from the
long-range \emph{interactions} embedded
in the loop entropy factor $\Omega(y-x)$.
In a probability profile, cooperativity appear as the
characteristic plateaus that indicate the tendency of blocks of basepairs
to ``melt as one''---being either all 0's or all 1's. Basepairs within
such a cooperative region are strongly
correlated. This aspect of cooperativity
is also prominent in a stitch profile,
where the block organization is shown
more specifically. For comparison, Fig.~\ref{prob}
shows a stitch profile and a probability profile of the same piece of DNA.
Stitches in the diagram are labeled with their peak volumes.
As can be seen, plateaus in the probability profile
can be identified with one or more stitches
where there are correspondences in position and between the peak volumes and
the plateau height relative to surrounding plateaus.

In addition to the strong correlations inside a cooperative region,
there are weaker correlations over longer distances.
For example, a certain loop at one site
can influence what loops that can exist at a distant site.
Information about such correlations across one or more stitch boundaries
may be derived from the alternative rows of stitches that
indicate the possible multi-loop conformations. However,
this has yet to be developed formally.

It can be expected that stitch profiles represent DNA cooperativity better
than probability profiles, when considering
the types of probabilities involved:
Correlations between basepairs $x_{i}$ and $x_{j}$ can be
formulated using conditional probabilities $p(x_{i}|x_{j})$.
Conditional probabilities can not be derived from a probability
profile $p(x_{i})$ alone, but they can be derived using block
probabilities $p(x_{i}\ldots x_{j})$.

Figure \ref{prob} also illustrates that some stitches
have a \emph{dead end}, that is,
a boundary that does not coincide with other stitches' boundaries.
A row of stitches can not be continued at a dead end.
Stitches with dead ends typically have low peak volumes.
Dead ends exist because the
continuation of a row splits up in several stitches that all have
peak volumes below the cutoff and are therefore not included.
It is possible to make stitch profiles without dead ends
by replacing the simple cutoff selection with
some other appropriate method.
\subsection{The parameters $D_{\text{max}}$ and $p_{\text{c}}$}
We have seen in
Fig.~\ref{lakesfig} how $D_{\text{max}}$ controls the lakes
found by the \textsc{maxdeep} algorithm. An effect of
increasing $D_{\text{max}}$ is to increase the widths of the
fluctuation bars (i.e. lakes) in a stitch profile. Another effect
has to do with the hierarchical merging of lakes: In Fig.~\ref{prob},
the fluctuation bars correspond to the sloping parts of the
probability profile. However, these sloping parts contain smaller plateaus.
Decreasing $D_{\text{max}}$ can reveal this internal structure, by splitting
a stitch into several stitches with smaller fluctuation bars.

\begin{figure}
 \includegraphics[width=8.6cm]{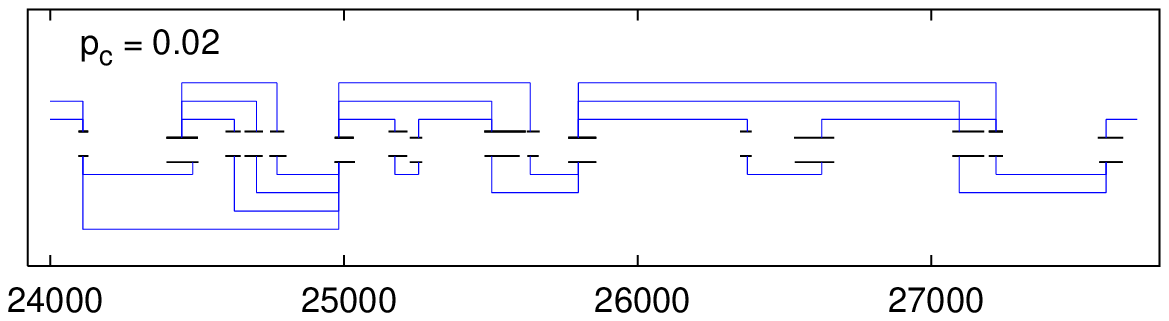}%

 \includegraphics[width=8.6cm]{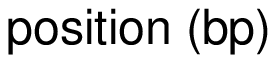}%

 \includegraphics[width=8.6cm]{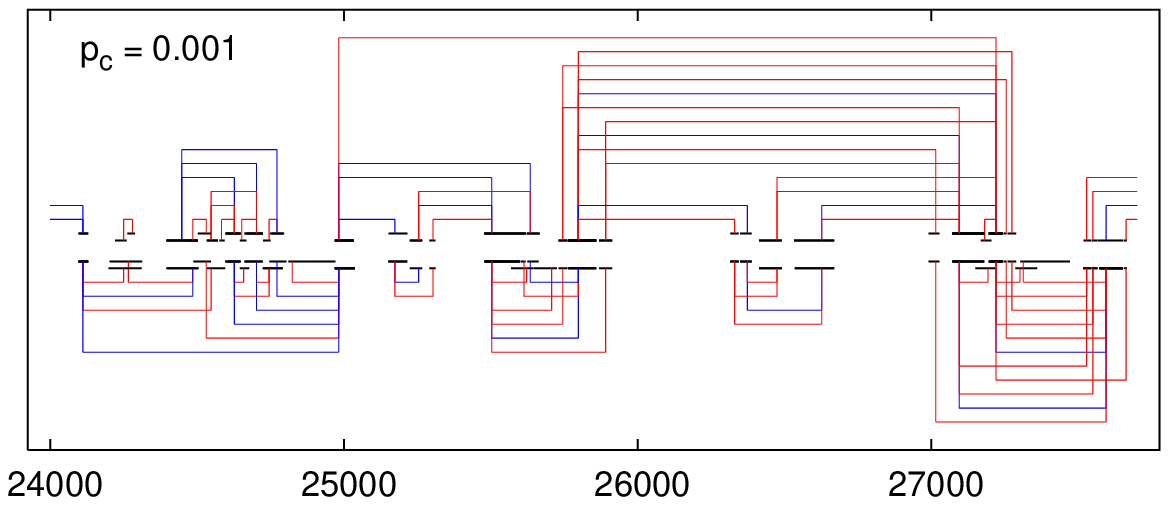}%

 \includegraphics[width=8.6cm]{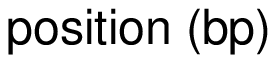}%
 \caption{
(Color online) Extra stitches are included when the cutoff
$p_{\text{c}}$ is lowered. The first stitch profile ($p_{\text{c}}=0.02$)
contains stitches with peak volumes $p_{\text{v}}\geq 0.02$.
The second stitch profile ($p_{\text{c}}=0.001$)
contains the same stitches as the first one (in blue), plus
an extra set of stitches with peak volumes
$0.001\leq p_{\text{v}}<0.02$ (in red).
The other parameters are $D_{\text{max}}=3$
and $T\simeq 80.6^{\circ }\text{C}$
and the sequence window is 24 kbp--27.7 kbp.
 \label{cutofffig}
}
 \end{figure}

Figure \ref{cutofffig}
illustrates the role of $p_{\text{c}}$. When $p_{\text{c}}$ is lowered,
an extra set of stitches is added to the stitch profile. $p_{\text{c}}$
does not modify the stitches as found
by the \textsc{maxdeep} algorithm, it only
controls how many of them are included in the diagram. The extra stitches
represent rare events (low probabilities), and they provide a more
finegrained picture like higher order terms in an expansion.
The number of stitches in a stitch profile
directly depends on $p_{\text{c}}$,
but it also depends on $D_{\text{max}}$ indirectly,
because the peak volumes depend on
the widths of lakes and frames.

The stitch profiles made using this article's methods
depend on the values of $D_{\text{max}}$ and $p_{\text{c}}$.
The choice of these values depends on what we want to see.
One could seek for alternative methods that are parameterfree or only use
one parameter by applying, for example, an optimization scheme
and some optimality criterion. But in my opinion,
reducing the number of parameters would
only hide the fact that the peak finding task
involves two different types of choice:
(1) lumping together related events into a peak
and (2) selecting what peaks to include.
The cutoff selection method using $p_{\text{c}}$ is simple to program,
requires only little computer power, and
can reuse data from a stitch profile
that has a lower cutoff.
Therefore, $p_{\text{c}}$ can be used in practice for ``finetuning''
by trying out different values iteratively.
In this way, we can make
a stitch profile with a certain total number of stitches.
Or a stitch profile with a certain maximum height of the stackings
of stitches in the diagram, which would limit the visual complexity
(cf.\ Fig.~\ref{cutofffig}).
\subsection{Ensemble representation}
\begin{figure}
 \includegraphics[width=8.6cm]{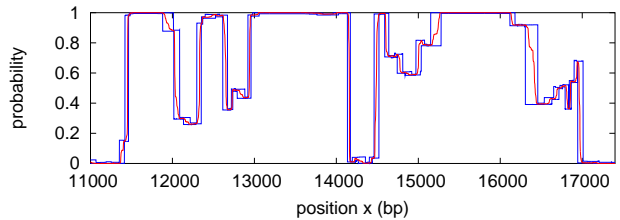}%
 \caption{
 (Color online)
 The probability profile $1-p_{\text{bp}}(x)$ of Fig.~\ref{prob}
 is plotted here (in red) with
 its upper and lower bounds (in blue),
 $1-p_{\text{low}}(x)$ and $1-p_{\text{up}}(x)$, obtained
 from the stitch profile.
 \label{bounds}
 }
\end{figure}

The alternative conformations represented by a stitch profile are few
in numbers compared to the total number $2^N$ of possible conformations.
But they may constitute a considerable fraction of the ensemble
in terms of statistical weight. Is this fraction big or small? A direct
answer would be obtained by comparing the total partition function $Z$
with the partition function restricted to the stitch profile conformations.
Instead, we take a graphical approach that relates peak volumes
to basepairing probabilities.
A stitch profile provides \emph{upper} and \emph{lower bounds}
of the corresponding probability profile. For example, the presence
of a loop stitch as in Fig.~\ref{row} implies that basepairs in
that region are melted with probability greater than the peak volume.
Or more precisely, $1-p_{\text{bp}}(x)\geq p_{\text{v}}(a,b)$ for
$L_{\text{R}}(a)<x<L_{\text{L}}(b)$. Stitches that overlap are
mutually exclusive, so we can sum over stitches to obtain
bounds as follows. Recall that an indicator function is defined as
\begin{equation}
I_{[i,j]}(x)=\left\{
                     \begin{array}{l}
                     1 \text{ for } x\in [i,j]\\
                     0 \text{ otherwise }.
 		     \end{array}
		\right.
\label{indicator}
\end{equation}
Then $p_{\text{low}}(x)\leq p_{\text{bp}}(x)\leq p_{\text{up}}(x)$, where
\begin{eqnarray}
p_{\text{up}}(x) &=&
1-\sum_{\text{left tail }a}I_{[1,L_{\text{L}}(a)-1]}(x)p_{\text{v}}(a)
\nonumber \\
 & & -\sum_{\text{loop }(a,b)}I_{[L_{\text{R}}(a)+1,L_{\text{L}}(b)-1]}(x)
 p_{\text{v}}(a,b)
\nonumber \\
 & &  -\sum_{\text{right tail }a}I_{[L_{\text{R}}(a)+1,N]}(x)p_{\text{v}}(a),
\label{lower}
\end{eqnarray}
and
\begin{eqnarray}
p_{\text{low}}(x) =
\sum_{\text{helix }(a,b)}I_{[L_{\text{R}}(a),L_{\text{L}}(b)]}(x)
p_{\text{v}}(a,b).
\label{upper}
\end{eqnarray}
Figure \ref{bounds}
shows the probability profile $1-p_{\text{bp}}(x)$
from Fig.~\ref{prob}, together with
its two bounds calculated using the stitch profile in Fig.~\ref{prob}
and Eqs.~(\ref{lower}) and (\ref{upper}).
The three curves are quite close. The two bounding curves consist of
vertical and horizontal lines because of the block nature of the stitch
profile. Given this constraint, they almost come as close as possible
to the probability profile. The probability profile
can apparently be reproduced from the stitch profile with only a small error.
This suggests that the conformations represented by the
stitch profile constitute a majority of the ensemble
in terms of statistical weight.
\subsection{Predicting metastable conformations}
A depth between zero and $D_{\text{max}}$ is associated
with each stitch in a stitch profile. A depth relates
to the landscape picture, but it actually characterizes
the probabilities of boundary positions:
The probability of a boundary in the fluctuation interval $L(a)$
to be located at $\beta a$ is $10^{D(a)}$ times greater than
the probability of being at $L_{\text{L}}(a)$ or $L_{\text{R}}(a)$.

The landscape picture of lakes confined by barriers
has been borrowed in this article for the purpose
of probability peak finding. This should
be distinguished from the usual purpose of describing the nonequilibrium
behavior of systems that are ``trapped'' in metastable states \cite{stein}.
Nevertheless, an idea of dynamics is implicit when talking about fluctuations.
Are the fluctuation bars related to actual fluctuations over time?

In a dynamics interpretation, a 1D lake $L(a)$
describes the ranges of the diffusion of a boundary
position on timescale $\tau \propto 10^{D(a)}$. A stitch profile would then
predict the ranges of fluctuations that
can be observed in an experiment during
time $\tau \propto 10^{D_{\text{max}}}$ with an empirical constant
of proportionality. However, this timescales interpretation
is preliminary until the rates of nucleation events creating new loops,
tails, and helices are accounted for.
If the depths are related to timescales, a possible application of
stitch profiles is to predict metastable conformations
that could play an important role in intracellular DNA or
other nonequilibrium situations.
Stitches that have large depths
(i.e. long-lived) and small peak volumes (i.e. rare)
are expected to indicate metastable conformations.
They can be easily found using a slightly modified cutoff selection method.
It is more difficult to detect such conformations
in a probability profile because of their low probabilities.
\subsection{Applications}
In conclusion, peaks in the block probabilities
[Eqs.~(\ref{ploop})--(\ref{phelix})]
can be found and represented as stitches in a stitch profile.
A stitch profile indicates the sizes and locations of loops,
tails and helical regions, their probabilities and ``depths'',
and how they fluctuate.
Multi-loop conformations can be derived from the
alternative rows of stitches, which may
show correlations between distant stitches.
A stitch profile thus predicts the conformations of partly melted DNA
and can account for a majority of the conformational ensemble
in terms of the basepairing probabilities.

Stitch profiles are motivated by the general idea that
a better prediction of DNA's conformational behavior
may contribute to a better understanding of DNA's functional
behavior, when there is a structure-function relationship.
Strand separation is at the heart of various processes
that occur in chromatin and chromosomes \cite{sumner}.
The question is, what role does the sequence dependence
of loop stabilities as predicted for DNA melting play
in biology? It is reasonable to believe that the low stability of
AT-rich regions is important in origins of replication \cite{sumner}
and in transcription initiation \cite{choi}.
Furthermore, bioinformatic evidence
points at more extensive and not yet explained correlations
in some genomes between the predicted melting properties
and the organization along the sequence of exons, introns,
and other genetic elements
\cite{Yeragenea,Yerageneb,Yeraproof,websidd,wada}.
There are different hypotheses. One view is that DNA mainly is
digital information storage and that such correlations is a
secondary effect reflecting, for example, varying compositions
of proteins \cite{poland2004}. Another view is that
DNA is also physical and that loop stabilities and/or other
sequence dependent biophysical properties of DNA
contribute actively in different biological mechanisms.
Some more or less speculative examples are: recombination
and crossover, sister chromatid adhesion, DNA-protein interactions,
and intron insertion \cite{exons}.
DNA conformational changes in a cell are not driven by
temperature changes, but rather by molecular forces and
interactions. Why then are DNA melting predictions relevant?
The Poland-Scheraga model deals with \emph{in vitro}
conditions that are far from the conditions in chromatin:
crowding \cite{ellis} is not accounted for in the loop entropy factor,
condensation and protein interactions are missing, and topology and chromosomal
geography is not accounted for. Fortunately, the correlations
found by Yeramian and others suggest a robustness of the predicted
melting properties. Stitch profiles may also apply to
intracellular DNA.

As mentioned in the Introduction, the development of stitch profiles
is also motivated by new single molecule techniques, in which some of the
predicted properties could be measured. For example, it would
be interesting to compare with measurements of
bubble sizes and their statistical weights \cite{zeng},
positions and stabilities of ``tails'' \cite{danilo}, and
bubble lifetimes \cite{altan}. As explained in the previous section,
it is an open question how the depths of stitches relate to
the lifetimes of the corresponding conformational features.

Stitch profiles may supplement the use of ordinary melting profiles
in the design and interpretation of \emph{in vitro} experiments
such as gel electrophoresis \cite{lerman87,steger} and in
the design of probes and primers for PCR and microarrays
\cite{halperin}. For example, stitch profiles
emphasize the ensemble aspect and may thereby predict some
features of gel experimental data.
For short DNAs, however, it is relevant to consider also
secondary structure, slippage and mismathes \cite{zukerhybrid,zhang-chen},
that are not accounted for in the Poland-Scheraga model.

A web server for computing stitch profiles has been made
available at http://stitchprofiles.uio.no \cite{narweb}.
\begin{acknowledgments}
I thank Edouard Yeramian, Einar R\o dland, Eivind Hovig, Enrico Carlon,
Fang Liu, Geir Ivar Jerstad and Tor-Kristian Jenssen.
\end{acknowledgments}

\bibliography{stitchqbio2}

\begin{thebibliography}{33}
\expandafter\ifx\csname natexlab\endcsname\relax\def\natexlab#1{#1}\fi
\expandafter\ifx\csname bibnamefont\endcsname\relax
  \def\bibnamefont#1{#1}\fi
\expandafter\ifx\csname bibfnamefont\endcsname\relax
  \def\bibfnamefont#1{#1}\fi
\expandafter\ifx\csname citenamefont\endcsname\relax
  \def\citenamefont#1{#1}\fi
\expandafter\ifx\csname url\endcsname\relax
  \def\url#1{\texttt{#1}}\fi
\expandafter\ifx\csname urlprefix\endcsname\relax\def\urlprefix{URL }\fi
\providecommand{\bibinfo}[2]{#2}
\providecommand{\eprint}[2][]{\url{#2}}

\bibitem[{\citenamefont{Sumner}(2003)}]{sumner}
\bibinfo{author}{\bibfnamefont{A.~T.} \bibnamefont{Sumner}},
  \emph{\bibinfo{title}{Chromosomes. Organization and function}}
  (\bibinfo{publisher}{Blackwell}, \bibinfo{address}{Oxford},
  \bibinfo{year}{2003}).

\bibitem[{\citenamefont{Poland}(1974)}]{poland}
\bibinfo{author}{\bibfnamefont{D.}~\bibnamefont{Poland}},
  \bibinfo{journal}{Biopolymers} \textbf{\bibinfo{volume}{13}},
  \bibinfo{pages}{1859} (\bibinfo{year}{1974}).

\bibitem[{\citenamefont{Yeramian and Jones}(2003)}]{Yerafizz}
\bibinfo{author}{\bibfnamefont{E.}~\bibnamefont{Yeramian}} \bibnamefont{and}
  \bibinfo{author}{\bibfnamefont{L.}~\bibnamefont{Jones}},
  \bibinfo{journal}{Nucleic Acids Res.} \textbf{\bibinfo{volume}{31}},
  \bibinfo{pages}{3843} (\bibinfo{year}{2003}).

\bibitem[{\citenamefont{Blake and Delcourt}(1998)}]{BD98}
\bibinfo{author}{\bibfnamefont{R.~D.} \bibnamefont{Blake}} \bibnamefont{and}
  \bibinfo{author}{\bibfnamefont{S.~G.} \bibnamefont{Delcourt}},
  \bibinfo{journal}{Nucleic Acids Res.} \textbf{\bibinfo{volume}{26}},
  \bibinfo{pages}{3323} (\bibinfo{year}{1998}).

\bibitem[{\citenamefont{Blake et~al.}(1999)\citenamefont{Blake, Bizzaro, Blake,
  Day, Delcourt, Knowles, Marx, and SantaLucia}}]{meltsim}
\bibinfo{author}{\bibfnamefont{R.~D.} \bibnamefont{Blake}},
  \bibinfo{author}{\bibfnamefont{J.~W.} \bibnamefont{Bizzaro}},
  \bibinfo{author}{\bibfnamefont{J.~D.} \bibnamefont{Blake}},
  \bibinfo{author}{\bibfnamefont{G.~R.} \bibnamefont{Day}},
  \bibinfo{author}{\bibfnamefont{S.~G.} \bibnamefont{Delcourt}},
  \bibinfo{author}{\bibfnamefont{J.}~\bibnamefont{Knowles}},
  \bibinfo{author}{\bibfnamefont{K.~A.} \bibnamefont{Marx}}, \bibnamefont{and}
  \bibinfo{author}{\bibfnamefont{J.}~\bibnamefont{SantaLucia}},
  \bibinfo{journal}{Bioinformatics} \textbf{\bibinfo{volume}{15}},
  \bibinfo{pages}{370} (\bibinfo{year}{1999}).

\bibitem[{\citenamefont{Altan-Bonnet et~al.}(2003)\citenamefont{Altan-Bonnet,
  Libchaber, and Krichevsky}}]{altan}
\bibinfo{author}{\bibfnamefont{G.}~\bibnamefont{Altan-Bonnet}},
  \bibinfo{author}{\bibfnamefont{A.}~\bibnamefont{Libchaber}},
  \bibnamefont{and}
  \bibinfo{author}{\bibfnamefont{O.}~\bibnamefont{Krichevsky}},
  \bibinfo{journal}{Phys. Rev. Lett.} \textbf{\bibinfo{volume}{90}},
  \bibinfo{pages}{138101} (\bibinfo{year}{2003}).

\bibitem[{\citenamefont{Danilowicz et~al.}(2003)\citenamefont{Danilowicz,
  Coljee, Bouzigues, Lubensky, Nelson, and Prentiss}}]{danilo}
\bibinfo{author}{\bibfnamefont{C.}~\bibnamefont{Danilowicz}},
  \bibinfo{author}{\bibfnamefont{V.~W.} \bibnamefont{Coljee}},
  \bibinfo{author}{\bibfnamefont{C.}~\bibnamefont{Bouzigues}},
  \bibinfo{author}{\bibfnamefont{D.~K.} \bibnamefont{Lubensky}},
  \bibinfo{author}{\bibfnamefont{D.~R.} \bibnamefont{Nelson}},
  \bibnamefont{and} \bibinfo{author}{\bibfnamefont{M.}~\bibnamefont{Prentiss}},
  \bibinfo{journal}{Proc. Natl. Acad. Sci. USA} \textbf{\bibinfo{volume}{100}},
  \bibinfo{pages}{1694} (\bibinfo{year}{2003}).

\bibitem[{\citenamefont{Zeng et~al.}(2003)\citenamefont{Zeng, Montrichok, and
  Zocchi}}]{zeng}
\bibinfo{author}{\bibfnamefont{Y.}~\bibnamefont{Zeng}},
  \bibinfo{author}{\bibfnamefont{A.}~\bibnamefont{Montrichok}},
  \bibnamefont{and} \bibinfo{author}{\bibfnamefont{G.}~\bibnamefont{Zocchi}},
  \bibinfo{journal}{Phys. Rev. Lett.} \textbf{\bibinfo{volume}{91}},
  \bibinfo{pages}{148101} (\bibinfo{year}{2003}).

\bibitem[{\citenamefont{Hanke and Metzler}(2003)}]{hanke}
\bibinfo{author}{\bibfnamefont{A.}~\bibnamefont{Hanke}} \bibnamefont{and}
  \bibinfo{author}{\bibfnamefont{R.}~\bibnamefont{Metzler}},
  \bibinfo{journal}{J. Phys. A} \textbf{\bibinfo{volume}{36}},
  \bibinfo{pages}{L473} (\bibinfo{year}{2003}).

\bibitem[{\citenamefont{Hwa et~al.}(2003)\citenamefont{Hwa, Marinari, Sneppen,
  and h.~Tang}}]{hwa}
\bibinfo{author}{\bibfnamefont{T.}~\bibnamefont{Hwa}},
  \bibinfo{author}{\bibfnamefont{E.}~\bibnamefont{Marinari}},
  \bibinfo{author}{\bibfnamefont{K.}~\bibnamefont{Sneppen}}, \bibnamefont{and}
  \bibinfo{author}{\bibfnamefont{L.}~\bibnamefont{h.~Tang}},
  \bibinfo{journal}{Proc. Natl. Acad. Sci. USA} \textbf{\bibinfo{volume}{100}},
  \bibinfo{pages}{4411} (\bibinfo{year}{2003}).

\bibitem[{\citenamefont{Peyrard}(2004)}]{peyrardreview}
\bibinfo{author}{\bibfnamefont{M.}~\bibnamefont{Peyrard}},
  \bibinfo{journal}{Nonlinearity} \textbf{\bibinfo{volume}{17}},
  \bibinfo{pages}{R1} (\bibinfo{year}{2004}).

\bibitem[{\citenamefont{T{\o}stesen et~al.}(2003)\citenamefont{T{\o}stesen,
  Liu, Jenssen, and Hovig}}]{bip2003}
\bibinfo{author}{\bibfnamefont{E.}~\bibnamefont{T{\o}stesen}},
  \bibinfo{author}{\bibfnamefont{F.}~\bibnamefont{Liu}},
  \bibinfo{author}{\bibfnamefont{T.-K.} \bibnamefont{Jenssen}},
  \bibnamefont{and} \bibinfo{author}{\bibfnamefont{E.}~\bibnamefont{Hovig}},
  \bibinfo{journal}{Biopolymers} \textbf{\bibinfo{volume}{70}},
  \bibinfo{pages}{364} (\bibinfo{year}{2003}).

\bibitem[{\citenamefont{Poland and Scheraga}(1970)}]{PS}
\bibinfo{author}{\bibfnamefont{D.}~\bibnamefont{Poland}} \bibnamefont{and}
  \bibinfo{author}{\bibfnamefont{H.~A.} \bibnamefont{Scheraga}},
  \emph{\bibinfo{title}{Theory of helix-coil transitions in biopolymers}}
  (\bibinfo{publisher}{Academic Press}, \bibinfo{address}{New York},
  \bibinfo{year}{1970}).

\bibitem[{\citenamefont{T{\o}stesen et~al.}(2001)\citenamefont{T{\o}stesen,
  Chen, and Dill}}]{TCD2001}
\bibinfo{author}{\bibfnamefont{E.}~\bibnamefont{T{\o}stesen}},
  \bibinfo{author}{\bibfnamefont{S.-J.} \bibnamefont{Chen}}, \bibnamefont{and}
  \bibinfo{author}{\bibfnamefont{K.~A.} \bibnamefont{Dill}},
  \bibinfo{journal}{J. Phys. Chem. B} \textbf{\bibinfo{volume}{105}},
  \bibinfo{pages}{1618} (\bibinfo{year}{2001}).

\bibitem[{\citenamefont{Stein and Newman}(1995)}]{stein}
\bibinfo{author}{\bibfnamefont{D.~L.} \bibnamefont{Stein}} \bibnamefont{and}
  \bibinfo{author}{\bibfnamefont{C.~M.} \bibnamefont{Newman}},
  \bibinfo{journal}{Phys. Rev. E} \textbf{\bibinfo{volume}{51}},
  \bibinfo{pages}{5228} (\bibinfo{year}{1995}).

\bibitem[{\citenamefont{Hoffmann and Sibani}(1988)}]{hoffmann}
\bibinfo{author}{\bibfnamefont{K.~H.} \bibnamefont{Hoffmann}} \bibnamefont{and}
  \bibinfo{author}{\bibfnamefont{P.}~\bibnamefont{Sibani}},
  \bibinfo{journal}{Phys. Rev. A} \textbf{\bibinfo{volume}{38}},
  \bibinfo{pages}{4261} (\bibinfo{year}{1988}).

\bibitem[{\citenamefont{Dauxois et~al.}(1993)\citenamefont{Dauxois, Peyrard,
  and Bishop}}]{DPB93}
\bibinfo{author}{\bibfnamefont{T.}~\bibnamefont{Dauxois}},
  \bibinfo{author}{\bibfnamefont{M.}~\bibnamefont{Peyrard}}, \bibnamefont{and}
  \bibinfo{author}{\bibfnamefont{A.~R.} \bibnamefont{Bishop}},
  \bibinfo{journal}{Phys. Rev. E} \textbf{\bibinfo{volume}{47}},
  \bibinfo{pages}{R44} (\bibinfo{year}{1993}).

\bibitem[{\citenamefont{Wada and Suyama}(1984)}]{wada}
\bibinfo{author}{\bibfnamefont{A.}~\bibnamefont{Wada}} \bibnamefont{and}
  \bibinfo{author}{\bibfnamefont{A.}~\bibnamefont{Suyama}},
  \bibinfo{journal}{J. Biomol. Str. \& Dyn.} \textbf{\bibinfo{volume}{2}},
  \bibinfo{pages}{573} (\bibinfo{year}{1984}).

\bibitem[{\citenamefont{Blossey and Carlon}(2003)}]{BC03}
\bibinfo{author}{\bibfnamefont{R.}~\bibnamefont{Blossey}} \bibnamefont{and}
  \bibinfo{author}{\bibfnamefont{E.}~\bibnamefont{Carlon}},
  \bibinfo{journal}{Phys. Rev. E} \textbf{\bibinfo{volume}{68}},
  \bibinfo{pages}{061911} (\bibinfo{year}{2003}).

\bibitem[{\citenamefont{Choi et~al.}(2004)\citenamefont{Choi, Kalosakas,
  Rasmussen, Hiromura, Bishop, and Usheva}}]{choi}
\bibinfo{author}{\bibfnamefont{C.~H.} \bibnamefont{Choi}},
  \bibinfo{author}{\bibfnamefont{G.}~\bibnamefont{Kalosakas}},
  \bibinfo{author}{\bibfnamefont{K.~{\O}.} \bibnamefont{Rasmussen}},
  \bibinfo{author}{\bibfnamefont{M.}~\bibnamefont{Hiromura}},
  \bibinfo{author}{\bibfnamefont{A.~R.} \bibnamefont{Bishop}},
  \bibnamefont{and} \bibinfo{author}{\bibfnamefont{A.}~\bibnamefont{Usheva}},
  \bibinfo{journal}{Nucleic Acids Res.} \textbf{\bibinfo{volume}{32}},
  \bibinfo{pages}{1584} (\bibinfo{year}{2004}).

\bibitem[{\citenamefont{Yeramian}(2000{\natexlab{a}})}]{Yeragenea}
\bibinfo{author}{\bibfnamefont{E.}~\bibnamefont{Yeramian}},
  \bibinfo{journal}{Gene} \textbf{\bibinfo{volume}{255}}, \bibinfo{pages}{139}
  (\bibinfo{year}{2000}{\natexlab{a}}).

\bibitem[{\citenamefont{Yeramian}(2000{\natexlab{b}})}]{Yerageneb}
\bibinfo{author}{\bibfnamefont{E.}~\bibnamefont{Yeramian}},
  \bibinfo{journal}{Gene} \textbf{\bibinfo{volume}{255}}, \bibinfo{pages}{151}
  (\bibinfo{year}{2000}{\natexlab{b}}).

\bibitem[{\citenamefont{Yeramian et~al.}(2002)\citenamefont{Yeramian, Bonnefoy,
  and Langsley}}]{Yeraproof}
\bibinfo{author}{\bibfnamefont{E.}~\bibnamefont{Yeramian}},
  \bibinfo{author}{\bibfnamefont{S.}~\bibnamefont{Bonnefoy}}, \bibnamefont{and}
  \bibinfo{author}{\bibfnamefont{G.}~\bibnamefont{Langsley}},
  \bibinfo{journal}{Bioinformatics} \textbf{\bibinfo{volume}{18}},
  \bibinfo{pages}{190} (\bibinfo{year}{2002}).

\bibitem[{\citenamefont{Bi and Benham}(2004)}]{websidd}
\bibinfo{author}{\bibfnamefont{C.}~\bibnamefont{Bi}} \bibnamefont{and}
  \bibinfo{author}{\bibfnamefont{C.~J.} \bibnamefont{Benham}},
  \bibinfo{journal}{Bioinformatics} \textbf{\bibinfo{volume}{20}},
  \bibinfo{pages}{1477} (\bibinfo{year}{2004}).

\bibitem[{\citenamefont{Poland}(2004)}]{poland2004}
\bibinfo{author}{\bibfnamefont{D.}~\bibnamefont{Poland}},
  \bibinfo{journal}{Biopolymers} \textbf{\bibinfo{volume}{73}},
  \bibinfo{pages}{216} (\bibinfo{year}{2004}).

\bibitem[{\citenamefont{Carlon et~al.}(2004)\citenamefont{Carlon, Malki, and
  Blossey}}]{exons}
\bibinfo{author}{\bibfnamefont{E.}~\bibnamefont{Carlon}},
  \bibinfo{author}{\bibfnamefont{M.~L.} \bibnamefont{Malki}}, \bibnamefont{and}
  \bibinfo{author}{\bibfnamefont{R.}~\bibnamefont{Blossey}}
  (\bibinfo{year}{2004}), \eprint{q-bio.BM/0409034}.

\bibitem[{\citenamefont{Ellis}(2001)}]{ellis}
\bibinfo{author}{\bibfnamefont{R.~J.} \bibnamefont{Ellis}},
  \bibinfo{journal}{Curr. Opin. Struct. Biol.} \textbf{\bibinfo{volume}{11}},
  \bibinfo{pages}{114} (\bibinfo{year}{2001}).

\bibitem[{\citenamefont{Lerman and Silverstein}(1987)}]{lerman87}
\bibinfo{author}{\bibfnamefont{L.~S.} \bibnamefont{Lerman}} \bibnamefont{and}
  \bibinfo{author}{\bibfnamefont{K.}~\bibnamefont{Silverstein}},
  \bibinfo{journal}{Methods Enzymol.} \textbf{\bibinfo{volume}{155}},
  \bibinfo{pages}{482} (\bibinfo{year}{1987}).

\bibitem[{\citenamefont{Steger}(1994)}]{steger}
\bibinfo{author}{\bibfnamefont{G.}~\bibnamefont{Steger}},
  \bibinfo{journal}{Nucleic Acids Res.} \textbf{\bibinfo{volume}{22}},
  \bibinfo{pages}{2760} (\bibinfo{year}{1994}).

\bibitem[{\citenamefont{Halperin et~al.}(2004)\citenamefont{Halperin, Buhot,
  and Zhulina}}]{halperin}
\bibinfo{author}{\bibfnamefont{A.}~\bibnamefont{Halperin}},
  \bibinfo{author}{\bibfnamefont{A.}~\bibnamefont{Buhot}}, \bibnamefont{and}
  \bibinfo{author}{\bibfnamefont{E.~B.} \bibnamefont{Zhulina}},
  \bibinfo{journal}{Biophys. J.} \textbf{\bibinfo{volume}{86}},
  \bibinfo{pages}{718} (\bibinfo{year}{2004}).

\bibitem[{\citenamefont{Dimitrov and Zuker}(2004)}]{zukerhybrid}
\bibinfo{author}{\bibfnamefont{R.~A.} \bibnamefont{Dimitrov}} \bibnamefont{and}
  \bibinfo{author}{\bibfnamefont{M.}~\bibnamefont{Zuker}},
  \bibinfo{journal}{Biophys. J.} \textbf{\bibinfo{volume}{87}},
  \bibinfo{pages}{215} (\bibinfo{year}{2004}).

\bibitem[{\citenamefont{Zhang and Chen}(2001)}]{zhang-chen}
\bibinfo{author}{\bibfnamefont{W.}~\bibnamefont{Zhang}} \bibnamefont{and}
  \bibinfo{author}{\bibfnamefont{S.-J.} \bibnamefont{Chen}},
  \bibinfo{journal}{J. Chem. Phys.} \textbf{\bibinfo{volume}{114}},
  \bibinfo{pages}{4253} (\bibinfo{year}{2001}).

\bibitem[{\citenamefont{T{\o}stesen et~al.}()\citenamefont{T{\o}stesen,
  Jerstad, and Hovig}}]{narweb}
\bibinfo{author}{\bibfnamefont{E.}~\bibnamefont{T{\o}stesen}},
  \bibinfo{author}{\bibfnamefont{G.~I.} \bibnamefont{Jerstad}},
  \bibnamefont{and} \bibinfo{author}{\bibfnamefont{E.}~\bibnamefont{Hovig}},
  \bibinfo{journal}{Nucleic Acids Res.} (to be published).

\end{thebibliography}

\end{document}